# On the nature of the residual meson-meson interaction from simulations with a QED$_{2+1}$ model [*]

H.R. Fiebig, O. Linsuain, [a] H. Markum and K. Rabitsch [b]

[a]Physics Department, F.I.U. - University Park, Miami, Florida 33199, U.S.A.

[b]Institut für Kernphysik, Technische Universität Wien, A-1040 Vienna, Austria

A potential between mesons is extracted from 4-point functions within lattice gauge theory taking 2+1 dimensional QED as an example. This theory possesses confinement and dynamical fermions. The resulting meson-meson potential has a short-ranged hard repulsive core and the expected dipole-dipole forces lead to attraction at intermediate distances. Sea quarks lead to a softer form of the total potential.

*1. Introduction.* The extraction of an effective interaction, or potential, between two composite hadrons from a quantum field theory is of fundamental importance. Here we have in mind the quark-gluon degrees of freedom at the subnuclear level. Those are thought to give rise to the nucleon-nucleon forces usually described by meson theory leading to the construction of the Bonn and Paris potentials [1].

In the last years lattice calculations with static quarks have demonstrated that the potential between two three-quark clusters is attractive [2]. A hard repulsive core of the potential, as suggested by experiments and their interpretation, could not be observed in the region where the two nucleons have relative distance close to zero. Beyond the static theory, employing dynamical quark propagators, correct antisymmetrisation and quark exchange become possible [3,4]. In this framework the mechanism which leads to a repulsive core will be investigated here analytically and numerically within lattice QED$_{2+1}$.

*2. Four-Point Correlation Functions.* We take the one-meson field as a product of staggered Grassmann fields $\chi$ and $\bar\chi$ with external flavours $u$ and $d$

$$\phi_{\vec x}(t) = \bar\chi_d(\vec x t)\chi_u(\vec x t) . \quad (1)$$

The meson-meson fields with relative distance $\vec r = \vec y - \vec x$ are then constructed by

$$\Phi_{\vec r}(t) = L^{-2}\sum_{\vec x}\sum_{\vec y}\delta_{\vec r, \vec y - \vec x}\phi_{\vec x}(t)\phi_{\vec y}(t) . \quad (2)$$

The information about the dynamics of the meson-meson system is contained in the 4-point time correlation matrix

$$C^{(4)}_{\vec r \vec r\,'}(t,t_0) = \langle\Phi^\dagger_{\vec r}(t)\Phi_{\vec r\,'}(t_0)\rangle - \langle\Phi^\dagger_{\vec r}(t)\rangle\langle\Phi_{\vec r\,'}(t_0)\rangle . \quad (3)$$

Working out the contractions between the Grassmann fields the following diagrammatic representation is obtained

$$C^{(4)} = C^{(4A)} + C^{(4B)} - C^{(4C)} - C^{(4D)} \quad (4)$$

$$= \;\|\,\| \;+\; \bowtie \;-\; \bowtie\!| \;-\; |\!\bowtie \;. \quad (5)$$

Each of the four contributions to the correlator comprises the exchange of gluons and sea quarks. For diagrams $C^{(4A)}$ and $C^{(4B)}$ those take place between the mesons, whereas diagrams $C^{(4C)}$ and $C^{(4D)}$ correspond also to interaction mediated by the exchange of valence quarks. Denoting the contracted Grassmann fields by the quark propagators $G_n, n=1\ldots 4$, we have for example

$$C^{(4A)} \sim \langle\phi^\dagger_{\vec y\,'}\phi^\dagger_{\vec x\,'}\phi_{\vec x}\phi_{\vec y}\rangle \overset{43\;21\;12\;34}{=} \langle G_1 G^*_2 G_3 G^*_4\rangle, \quad (6)$$

where $\sim$ stands for the sums with normalisation factors that carry over from (2).

In order to define an effective meson-meson interaction it is crucial that the noninteracting components in $C^{(4)}$ are isolated [4]. To the corresponding expression, $\bar C^{(4)} = \bar C^{(4A)} + \bar C^{(4B)}$, only the direct channels contribute. This respects the boson symmetry on the meson-field level and constitutes the free meson-meson correlator with the *relative* interaction switched off.

---

[*]Supported in part by NSF grant PHY-9409195, by FWF project P10468-PHY and by CEBAF.

*3. Effective Interaction from Time Evolution.* The deviation of $C^{(4)}$ from $\bar{C}^{(4)}$ contains the residual effective meson-meson interaction. From analogy with quantum mechanical perturbation theory

$$\hat{C}(t-t_0) = e^{-\hat{H}_I(t-t_0)}, \qquad (7)$$

we define an effective meson-meson interaction as

$$\mathcal{H}_I = -\left.\frac{\partial \mathcal{C}}{\partial t}\right|_{t=t_0}, \text{ with } \mathcal{C} = \bar{C}^{(4)-\frac{1}{2}} C^{(4)} \bar{C}^{(4)-\frac{1}{2}}. \quad (8)$$

This definition is meant as a relation between matrices and holds independently of the basis. It can be tested both in momentum space [5] and in coordinate space. The above equations build a bridge between the quantum field theoretical correlation functions on the lattice and an effective quantum mechanical Hamiltonian.

*4. Hard Core from Adiabatic Approximation.* An analysis of the coordinate-space matrix elements of the correlators $C^{(4)}$ and $\bar{C}^{(4)}$ reveals an interesting perspective on the repulsive core. From $C^{(4)}$ expressed in terms of the fermion propagators it is easy to see *analytically* that

$$C^{(4)}_{\vec{r}\vec{r}'}(t,t_0) \equiv 0, \quad \text{if} \quad \vec{r}=0 \text{ or } \vec{r}'=0. \quad (9)$$

In our simulation we choose the relative distance between the mesons to be the same at the initial and final times of the propagation, $\vec{r} = \vec{r}'$. In the spirit of the Born-Oppenheimer approximation it is assumed that the interaction proceeds faster than the motion between the mesons. This seems to be justified because the quark mass $m_F = 0.1$ is relatively large. In this sense we replace the sum over all eigenstates in the spectral representation of the correlation function by an average term with an effective energy $W(\vec{r})$

$$\begin{aligned} C^{(4)}_{\vec{r}\vec{r}}(t,t_0) &= \sum_n |\langle \vec{r}|n\rangle|^2 e^{-E_n(t-t_0)} \\ &\simeq c(\vec{r}) e^{-W(\vec{r})(t-t_0)}. \end{aligned} \quad (10)$$

To the extent that the energy $W(\vec{r})$ of the meson-meson system at fixed relative distance $\vec{r}$ can be extracted from the large-$t$ behaviour of the diagonal elements we may conclude from (9)–(10) that $W(\vec{r}=0) = +\infty$, provided the strength factor $c(\vec{r})$ is nonzero at $\vec{r}=0$. In this case the effective interaction possesses a hard repulsive core. It is due to the anticommuting nature of the constituent fermion fields leading to Pauli repulsion.

*5. Results.* The simulation was performed on an $L^2 \times T = 24^2 \times 32$ periodic lattice with the compact $U(1)$ Wilson action at $\beta = 1.5$ and staggered fermions with flavour number $N_F = 0$ and $N_F = 2$, both with dynamical quark mass $m_F = 0.1$. We used 64 independent gauge field configurations generated with the molecular dynamics algorithm [6]. Fermion propagators were computed with 32 and 16 random sources [7] in the quenched case and in full QED$_{2+1}$, respectively.

The potentials were extracted from cosh-fits to the correlator ratios $C^{(4)}/\bar{C}^{(4)}$ for $t = 15 \ldots 19$ in periodic time. The first definition (7)–(8) represents a matrix and leads to a nonlocal Hamilton operator $\mathcal{H}_I(\vec{r},\vec{r}')$ where $\vec{r},\vec{r}'$ is the separation between the mesons. Since computation of the entire 4-point correlation matrix is very time-consuming we only obtained the diagonal elements. This also allows a direct comparison with $\mathcal{W}_I(\vec{r}) = W(\vec{r}) - 2m$ from the second definition (10), where $2m$ is the mass of the noninteracting two-meson system.

We begin the description of the results of our simulation with the quenched theory, $N_F = 0$. In fig. 1 we display data of the correlation function $\mathcal{C}_{\vec{r}\vec{r}}$ for the direct channels (A+B), the quark-exchange channels (C+D) and for their combination (A+B)−(C+D). One observes a zig-zag behaviour for even and odd distances $r$. In the staggered scheme the meson correlator exhibits an oscillating term $\propto (-1)^t e^{-m't}$ with $m'$ almost degenerate with the ground state, $m' \approx m$. On a euclidean lattice there is a corresponding term in the space directions. This feature enters into the meson-meson correlator. For on-axis points $\vec{r}$ the large-$t$ behaviour of the correlator is

$$\mathcal{C}_{\vec{r}\vec{r}} \approx c^{(+)}(\vec{r}) e^{-\mathcal{W}_I^{(+)}(\vec{r})t} + (-)^r c^{(-)}(\vec{r}) e^{-\mathcal{W}_I^{(-)}(\vec{r})t}.$$

If one is interested in the meson-meson interaction between ground states only it is neces-sary to separate $c^{(\pm)}$. Independent 3-parameter fits to even and odd sets of $r$ were made using $f(r) = \alpha_1 e^{-\alpha_2 r} + \alpha_3$. Remarkably, the pairs of interpolated curves $c^{(\pm)}$ in fig. 1 exhibit very similar shapes (aside from different clustering). This indicates that the m–m and m′–m′ interactions are similar, thus justifying the use of the staggered scheme.

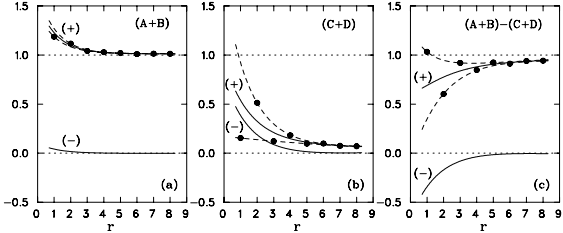

Figure 1. Raw data of correlator ratios $\mathcal{C}$ (symbols) and coefficients $c^{(\pm)}$ from an even-odd analysis as a function of relative distance $r$ at $t = 17$. Comparison of the (a) direct, (b) quark-exchange and the (c) combined channels.

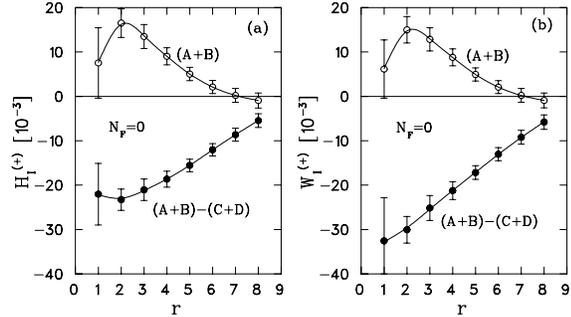

Figure 2. Meson-meson potentials for the direct channels (A+B) and all four graphs (A+B)–(C+D) extracted from interpolated correlators. The result (a) $\mathcal{H}_I(r)$ from the effective time evolution is compared with (b) $\mathcal{W}_I(r)$ of the adiabatic approximation. In both cases the direct channels lead to a repulsive potential barier. The total potential comes out attractive. Curves are guide to the eye and error bars result from a jack-knife analysis.

By repeating the above procedure for fixed $t$ and subsequent cosh-fits in $t$-direcïon we reconstructed the correlator ratios for the ground state meson-meson system. The results for $\mathcal{H}_I^{(+)}(r)$ are presented in fig. 2a and those for $\mathcal{W}_I^{(+)}(r)$ in fig. 2b. The (−) data are lost in noise. Both direct diagrams (A+B) correspond to a repulsive potential barier. The quark-exchange diagrams (C+D) yield similar behaviour (not shown) except their contributions enter with a minus sign into the total correlations (4),(5). This gives an attractive total potential as displayed in fig. 2.

We now consider correlations for all available distances $\vec{r}$ on the axes and diagonals. In fig. 3 the corresponding potentials of the direct diagrams (A+B) and all four diagrams (A+B)–(C+D) are shown. The direct channels turn out to be repulsive as in fig. 2. For both total potentials $\mathcal{H}_I(r)$ and $\mathcal{W}_I(r)$ there is now some evidence for a repulsive core, as discussed earlier, followed by attraction at medium distances. The shapes emerge as a subtle interplay between all graphs (A+B) and (C+D).

The Pauli principle leads us to expect a hard repulsive core. Our data show that its range is short. This is in contrast to a complementary simulation in a truncated momentum basis [4,5]. Therefore, the actual dynamical size of one meson on the lattice is of interest. Consider the fields

$$\phi_{\vec{\rho}}(t) = L^{-2} \sum_{\vec{x},\vec{y}} \delta_{\vec{\rho},\vec{x}-\vec{y}} \bar{\chi}_d(\vec{y}t) U_{\vec{\rho}}(\vec{y}t,\vec{x}t) \chi_u(\vec{x}t),$$

where the product of link variables $U_{\vec{\rho}}(\vec{y}t,\vec{x}t)$ along the quark-antiquark separation $\vec{\rho}$ ensures gauge invariance of the extended meson operator. The corresponding correlation matrix $C^{(2)}_{\vec{\rho},\vec{\rho}'}(t,t_0)$ was computed for $\rho,\rho' = 0,1,2$. Reduced matrices of $C^{(2)}$ for all relevant irreducible representations of the lattice symmetry group $O(2,\mathcal{Z})$ were obtained. It turned out that the matrix elements with $\rho,\rho' \neq 0$ were numerically zero. This indicates that, for the current model, the meson actually stays small ($\leq a$) during its propagation. With regard to the meson-meson interaction, this explains the narrow hard core.

Simulations for $N_F = 2$ flavours also have been performed. Using only 16 random sources the data are considerably more noisy. For ease of comparison, we rescale in fig. 4 $\mathcal{H}_I$ and $\mathcal{W}_I$ by a factor equal to the mass ratio $2m(N_F = 0)/2m(N_F = 2) = 1.031(9)/0.63(1) = 1.65$. For the total potentials there seem no qualitative changes compared to the pure gluonic case, except that sea quarks apparently lead to a softer behaviour. The shapes of $\mathcal{W}_I$ again resemble $\mathcal{H}_I$. The potentials from the direct channels are small.

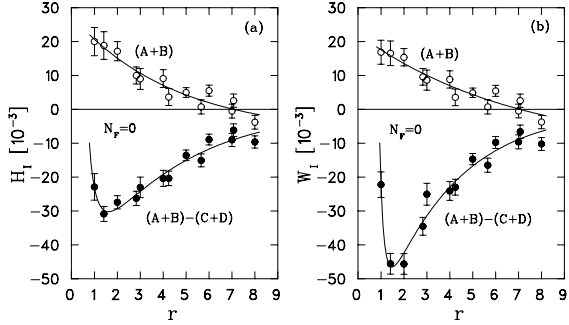

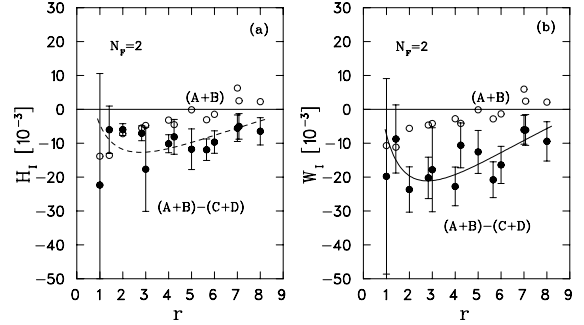

Figure 3. Meson-meson potentials as in fig. 2, but from raw correlators. The direct channels are repulsive again. Here both total potentials hint at a repulsive core at short range.

Figure 4. Meson-meson potentials as in fig. 2, extracted from full QED$_{2+1}$. The direct channels (open symbols without error bars) are small. The total potentials resemble those of the pure gauge case.

6. *Conclusion.* We have proposed two possibilities to extract an effective interaction between two composite particles. We found for QED$_{2+1}$ that gluon exchange and valence-quark exchange are equally important. The definite shape of the potential might depend on sea quark effects. In the model used mesons feel a short-ranged hard-core potential due the Pauli principle. The medium-range residual forces between colour-neutral quark-clusters are attractive. It should be noted that a computation of meson-meson scattering phase shifts based on Lüscher's proposal [8] indicates the same behaviour of the potential [9] as do a momentum-space study [4,5] and a different coordinate-space approach [3]. The correct antisymmetrisation of the meson-meson correlator was important for the emergence of a hard core. In contrast to $U(1)$, the nonabelian colour structure of $SU(3)$ will not necessarily lead to a vanishing correlator for nucleon-nucleon distance zero. Preliminary studies seem to indicate attraction. Calculations with light valence quarks might help to understand the hard-core problem of the nucleon-nucleon interaction.

## REFERENCES

1. K. Holinde, Phys. Rep. 68 (1981) 121; M. Lacombe, B. Loiseau, R. Vinh Mau, J. Côté, P. Pirès and R. de Tourreil, Phys. Rev. C23 (1981) 2405.
2. K. Rabitsch, H. Markum and W. Sakuler, Phys. Lett. B318 (1993) 507; A.M. Green, C. Michael and J.E. Paton, Nucl. Phys. A554 (1993) 701.
3. J.D. Canosa and H.R. Fiebig, Nucl. Phys. B (Proc. Suppl.) 34 (1994) 561.
4. J.D. Canosa, H.R. Fiebig and H. Markum, Nucl. Phys. B (Proc. Suppl.) 42 (1995) 657.
5. J.D. Canosa and H.R. Fiebig, F.I.U. Preprint, Miami, 1995, hep-lat/9508018.
6. S. Gottlieb, W. Liu, D. Toussaint, R.L. Renken and R.L. Sugar, Phys. Rev. D35 (1987) 2531.
7. R. Scalettar, D. Scalopino and R. Sugar, Phys. Rev. B34 (1986) 7911.
8. M. Lüscher, Nucl. Phys. B354 (1991) 531.
9. H.R. Fiebig, R.M. Woloshyn and A. Dominguez, Nucl. Phys. B418 (1994) 649.